\documentclass[letterpaper]{article} 
\usepackage{aaai2026}  
\usepackage{times}  
\usepackage{helvet}  
\usepackage{courier}  
\usepackage[hyphens]{url}  
\usepackage{xcolor}
\usepackage{xspace}
\usepackage{graphicx} 
\usepackage{natbib}  
\usepackage{caption} 
\usepackage{algorithm}
\usepackage{amsmath}
\usepackage{amssymb}
\usepackage{algpseudocode}
\usepackage{booktabs}
\usepackage{multirow}
\usepackage{amsmath}
\usepackage{tabularx}
\usepackage{tcolorbox}
\usepackage{graphicx}

\usepackage{color}
\usepackage{colortbl}
\usepackage{xcolor}
\usepackage[dvipsnames]{xcolor}
\definecolor{myorange1}{RGB}{255, 0, 0}
\definecolor{mygray}{gray}{.9}
\usepackage{cuted}

\usepackage{newfloat}
\usepackage{listings}
\DeclareCaptionStyle{ruled}{labelfont=normalfont,labelsep=colon,strut=off} 
\floatstyle{ruled}
\newfloat{listing}{tb}{lst}{}
\floatname{listing}{Listing}
\newcommand{\methodName}{GazeInterpreter\xspace}

\title{\methodName: Parsing Eye Gaze to Generate Eye-Body-Coordinated Narrations}
\author{
    Qing Chang$^1$\equalcontrib,
    Zhiming Hu$^{1,2}$\equalcontrib \\
}
\affiliations{
    $^1$The Hong Kong University of Science and Technology (Guangzhou), China\\
    $^2$The Hong Kong University of Science and Technology, China\\

    changqinghs@163.com, zhiminghu@hkust-gz.edu.cn
}

\usepackage{bibentry}

\def\shownotes{1}

\ifnum\shownotes=1
\newcommand\zhiming[1]{\textcolor{cyan}{Zhiming: #1}}
\newcommand\qing[1]{\textcolor{green}{Qing: #1}}

\else 
\newcommand\zhiming[1]{}
\newcommand\qing[1]{}

\newcommand\todo[1]{}
\fi

\begin{document}

\twocolumn[{%
\renewcommand\twocolumn[1][]{#1}%
\maketitle

\begin{center}
    \centering
    \captionsetup{type=figure}     
    \vspace{-5px}
    \includegraphics[width=\textwidth]{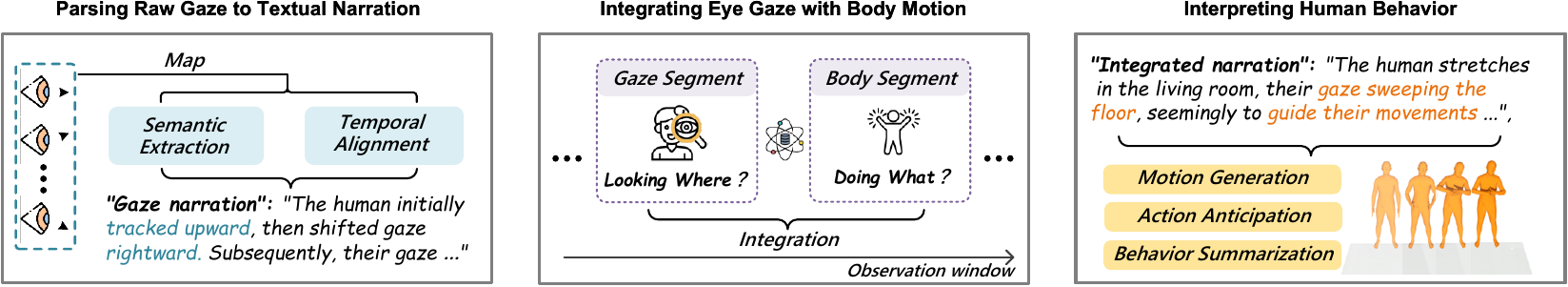}
    \caption{\methodName is a novel method for interpreting gaze behavior that first parses raw gaze signals to a textual narration (Left) and then integrates gaze with body motion to generate eye-body-coordinated narration (Middle). The integrated narration provides a superior basis for interpreting human behavior (Right). }
    \vspace{9px}
    \label{fig:motivation}
\end{center}%

}]
\begin{abstract}
Comprehensively interpreting human behavior is a core challenge in human-aware artificial intelligence.
However, prior works typically focused on body behavior, neglecting the crucial role of eye gaze and its synergy with body motion.
We present \textit{\methodName} -- a novel large language model-based (LLM-based) approach that parses eye gaze data to generate eye-body-coordinated narrations.
Specifically, our method features 1) a \textit{symbolic gaze parser} that translates raw gaze signals into symbolic gaze events; 2) a \textit{hierarchical structure} that first uses an LLM to generate eye gaze narration at semantic level and then integrates gaze with body motion within the same observation window to produce integrated narration; and 3) a \textit{self-correcting loop} that iteratively refines the \textit{modality match}, \textit{temporal coherence}, and \textit{completeness} of the integrated narration.
This hierarchical and iterative processing can effectively align physical values and semantic text in the temporal and spatial domains.
We validated the effectiveness of our eye-body-coordinated narrations on the text-driven motion generation task in the large-scale Nymeria benchmark.
Moreover, we report significant performance improvements for the sample downstream tasks of action anticipation and behavior summarization.
Taken together, these results reveal the significant potential of parsing eye gaze to interpret human behavior and open up a new direction for human behavior understanding. 
\end{abstract}

\begin{links}
\small
\link{Code}{https://github.com/EvergreenChang/GazeInterpreter}
\end{links}

\section{Introduction}
Comprehensively interpreting human behavior is a foundational challenge in human-aware artificial intelligence.
A robust understanding of human behavior underpins many critical applications, including human motion generation~\cite{zhang2022motiondiffuse, yan24gazemodiff}, human intention recognition~\cite{hu2022ehtask, belardinelli2022intention}, proactive action anticipation~\cite{hu24hoimotion, hu24gazemotion}, and efficient behavior summarization~\cite{zhang25summact}. 
With the recent success of large language models (LLMs), researchers have begun leveraging them to generate natural language explanations of human behavior, making notable progress in interpreting body motion~\cite{jiang2023motiongpt, chen2024motionllm}.

However, a crucial modality is largely overlooked in this new paradigm: human eye gaze.
As a powerful non-verbal cue, eye gaze is not only a direct window into human intention~\cite{hu2022ehtask, belardinelli2022intention} but is also intrinsically correlated with body motion~\cite{hu24pose2gaze, sidenmark2019eye}. For instance, when a person intends to grasp a cup, their eyes typically fixate on it just before or during the arm's movement. Despite this, prior works~\cite{kong2022human, chang2025spatial} have predominantly focused on interpreting body behavior in isolation, neglecting the potential information conveyed by eye gaze and its synergistic relationship with body motion. This omission results in a significant gap, leaving interpretations of human behavior incomplete and less robust.

To fill this gap, we present \textit{\methodName}~-- a novel LLM-based framework that parses eye gaze data to generate comprehensive eye-body-coordinated narrations, integrating human potential intentions and fine-grained motion features.
Specifically, we first introduce a \textit{symbolic gaze parser} that converts raw eye gaze signals into symbolic gaze events that serve as input to the LLM.
Then, a \textit{hierarchical structure} composed of multiple LLMs generates eye gaze narrations from gaze events, and these narrations are semantically integrated with body motion narrations, producing eye-body-coordinated narrations.
To ensure alignment between narrations and reality, we further present a \textit{self-correcting loop} that iteratively refines the \textit{continuity} of the gaze narrations and the \textit{modality match}, \textit{temporal coherence}, and \textit{completeness} of the integrated narrations by an LLM-driven evaluation-feedback mechanism.

We extensively evaluate our method for text-driven motion generation on the large-scale, in-the-wild
Nymeria benchmark~\cite{ma2024nymeria}. Experiments show that our eye-body-coordinated narrations lead to superior performance for generating motions. 
Furthermore, we validate the effectiveness of our narrations on representative downstream tasks of action anticipation and behavior summarization, demonstrating the advantages of our approach in comprehensive human behavior interpretation.

The contributions of our work are three-fold:
\begin{itemize}
    \item We propose~\textit{\methodName}~-- a novel LLM-based framework for interpreting gaze behavior that features a symbolic gaze parser to convert raw gaze signal, a hierarchical structure to integrate gaze with body motion, and a self-correcting loop for refinement. 
    \item We conduct extensive experiments on the large-scale Nymeria benchmark and demonstrate that our narrations can significantly improve performance in text-driven motion generation.
    \item We demonstrate the broad applicability of our method by significantly enhancing performance on the sample downstream tasks of action anticipation and behavior summarization.  
\end{itemize}

\section{Related Work}

\subsection{Human Behavior Interpreting}
Comprehensively interpreting human behavior is a crucial topic in the areas of human-centered computing and human-aware artificial intelligence.
Earlier works typically focused on rule-based techniques to interpret human behavior~\cite{pons2014posebits, delmas2022posescript}.
For example, Pons-Moll et al. manually set rules on \textit{joint distance}, \textit{articulation angle}, and \textit{relative position} to describe relationships between body parts~\cite{pons2014posebits} while Delmas et al. defined a set of rules on the 3D keypoints to generate the description of full-body pose~\cite{delmas2022posescript}.
Recently, with the great success of LLMs, many researchers have started to interpret human behavior directly using LLMs~\cite{jiang2023motiongpt, chen2024motionllm}.
Specifically, Jiang et al. proposed a uniform motion-language generative pre-trained model to link human body motion with natural language~\cite{jiang2023motiongpt} while Chen et al. projected human body motion and video data into the linguistic space to generate better narrations of human body behavior~\cite{chen2024motionllm}.
However, prior works mainly focused on interpreting human body behavior, neglecting the human eye gaze.
To fill this gap, in this work, we extract semantics from raw gaze signals and integrate them with body motion narrations.

\subsection{Eye Gaze Analysis}
Analyzing human gaze behavior has been a popular topic in the area of vision research for decades~\cite{yarbus1967eye, itti1998model, chen2024gazexplain}.
Prior works typically focused on hand-crafted statistical indicators of gaze behavior such as gaze velocity~\cite{hu2019sgaze, kothari2020gaze}, gaze distribution~\cite{sitzmann2018saliency, hu2019sgaze, hadnett2019effect}, saccade amplitude~\cite{hu2019sgaze, hadnett2019effect, hu2020dgaze}, fixation number~\cite{coutrot2018scanpath, hu2022ehtask}, fixation duration~\cite{hadnett2019effect, kothari2020gaze, hu2022ehtask}, fixation dispersion~\cite{coutrot2018scanpath, hu2022ehtask}, and fixation clusters~\cite{coutrot2018scanpath, hu2021fixationnet}.
However, these hand-crafted statistical indicators are cumbersome to compute, provide only limited information, and lack explainability.
In stark contrast, in this work, we directly convert raw gaze signals into natural language narrations that can effectively improve the understanding of eye gaze behavior.

\subsection{Eye-Body Coordination}
Many researchers have investigated the coordination of human eye gaze and their body movements.
Specifically, Kothari et al. focused on real-world scenarios and discovered the coordinated patterns of eye and head movements during daily activities~\cite{kothari2020gaze} while Hu et al. examined virtual environments and revealed the eye-head coordination during free-viewing and task-oriented situations~\cite{hu2019sgaze,hu2020dgaze,hu2021fixationnet,hu25hoigaze}.
Sidenmark et al. analyzed the gaze shift process in virtual reality and identified the coordination of eye, head, and torso movements~\cite{sidenmark2019eye}.
Hu et al. further revealed the strong correlation between eye gaze and human full-body movements in various daily activities~\cite{hu24pose2gaze}.
Inspired by the close link between eye gaze and body movements, in this work, we integrate eye gaze descriptions with body motion narrations to improve human behavior understanding.\looseness=-1
\section{Method}
\begin{figure*}[!t]
    \centering
    \includegraphics[width=.95\linewidth]{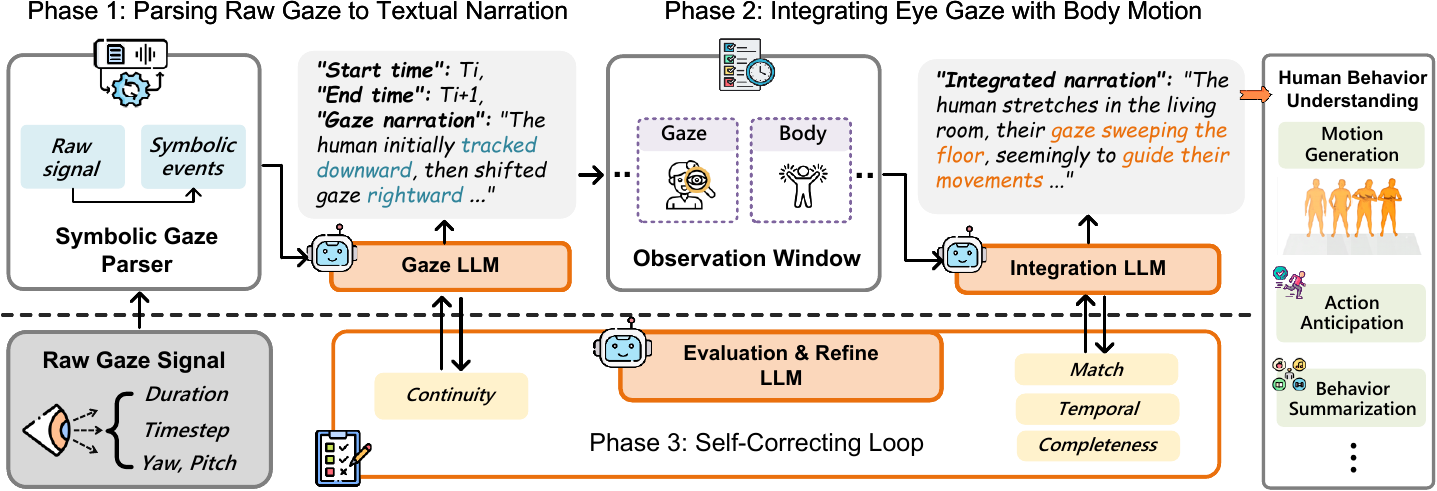}
   \caption{Architecture of \methodName. Our method first uses a symbolic gaze parser to convert raw gaze signals into symbolic events and then employs an LLM to generate textual narration (Phase 1), which is further integrated with body motion narration in an observation window to produce the eye-body-coordinated narration (Phase 2). A self-correcting loop is applied to iteratively refine both the gaze and integrated narration, ensuring feature alignment across different dimensions (Phase 3).
}
    \label{fig:pipeline}
\end{figure*}


\subsection{Problem Formulation}
We define interpreting eye gaze behavior as the task of generating comprehensive eye-body-coordinated narrations from eye gaze signals and body motion narrations.
For a given time segment $i$, we define the input as a multi-modal tuple $O_i = (S_i^{\text{g}}, S_i^{\text{m}})$, where $S_i^{\text{g}} \in \mathbb{R}^{N_g \times 2}$ represents the raw, continuous gaze signal, composed of $N_g$ samples of yaw and pitch coordinates. Concurrently, $S_i^{\text{m}} = (m_1, \dots, m_{N_m})$ is a sequence of $N_m$ discrete atomic body motion narrations from a predefined set $\mathcal{M}$.
Our objective is to learn a mapping $\mathcal{F}: (\mathbb{R}^{N_{g} \times 2}, \mathcal{M}^{N_{m}}) \to \mathcal{T}$, which generates a textual narration $\hat{T}_i \in \mathcal{T}$ that not only captures overt eye-body actions but also facilitates human behavior understanding.

A fundamental challenge in this task is to bridge the semantic gap between low-level, numerical gaze sensor data and the high-level, abstract concepts needed for comprehensive behavior understanding.
The core research question is: \textit{How can we reliably transform noisy, continuous gaze signals $S^{\text{g}}$ into a structured, semantic representation that facilitates faithful multi-modal reasoning?}\looseness=-1

While recent approaches have explored direct numerical-to-text conversion using LLMs~\cite{wang2024can} or employed auxiliary techniques like contrastive learning~\cite{chang2024hybrid}, such end-to-end methods remain susceptible to common pitfalls, including factual hallucination and a failure to robustly ground the generated text in the source signals~\cite{xu2025distinguishing}. To address these limitations, we propose to first abstract low-level raw gaze data into an intermediate symbolic vocabulary of gaze events, providing a reliable basis for motion fusion. This numerical-to-symbolic conversion contributes to the generation of more robust behavioral descriptions.\looseness=-1

\subsection{Framework Overview}
To address the formulated problem, we propose \textit{\methodName}~-- a novel hierarchical coarse-to-fine LLM-based method that transforms raw low-level gaze signals into high-level, interpretable narrations through three phases: \textit{parsing raw gaze to textual narration}, \textit{integrating eye gaze with body motion}, and \textit{self-correcting loop}, as depicted in Fig.~\ref{fig:pipeline}.
In the first phase, our method parses the raw gaze signal into a sequence of structured symbolic gaze events and then applies an LLM to generate texture narrations of these structured gaze events.
Benefiting from the results of the first phase, we can align eye gaze narrations with atomic body motion narrations in the semantic space during the next phase and synthesize eye-body-coordinated narrations.
In the third phase, our method applies a self-correcting loop to iteratively refine both the gaze narrations and the integrated narrations based on explicit quality criteria.

This decomposed, hierarchical coarse-to-fine architecture is central to our contribution. It ensures that the generated narrations are not only robustly grounded in observable sensor data but are also contextually holistic. 
By design, this approach overcomes the factual inconsistency and local incoherence that often plague monolithic, end-to-end models~\cite{lee2023well,yan2023efficient}, leading to more faithful and useful representations of human behavior.


\begin{table*}[ht]
    \centering
    \resizebox{\linewidth}{!}{
    \begin{tabular}{ccll}
        \toprule
        \multicolumn{1}{c}{\textbf{Type}} & \multicolumn{1}{c}{\textbf{Dimesion}} & \multicolumn{1}{c}{\textbf{High}} & \multicolumn{1}{c}{\textbf{Low}} \\
        \midrule
        Gaze & Continuity & 5: Perfect gaze transitions with natural flow. &
        0: Contains abrupt, illogical, or disjointed event descriptions. \\
       \midrule
        \multirow{3}{*}{Integrated} & Match & 5: Mutually supportive integration of modalities. &
        0: Modalities are disconnected, redundant, or contradictory. \\
        
         & Temporal & 5: Clear, logical, chronological progression. &
        0: Lacks a discernible temporal structure or causal flow.  \\
      
         & Completeness & 5: Fully includes all key elements and actions. &
        0: Essential information or key behavioral events are omitted. \\
        \bottomrule
    \end{tabular}
    }
    \caption{Scoring rubric for the multi-dimensional narration evaluation.}
    \label{tab:eval_criteria}
\end{table*}

\subsection{Phase 1: Parsing Raw Gaze to Textual Narration}\label{sec:symbolic}
The first phase of \methodName is \textit{parsing raw gaze to textual narration}, designed to bridge the gap between raw, continuous sensor readings and a discrete, semantic representation.
Directly interpreting noisy numerical data with LLMs poses significant grounding challenges, often leading to factual inconsistencies~\cite{bommasani2021opportunities,lee2023well}. 
To circumvent this, we adopt a two-step approach that decomposes the problem into deterministic symbolic parsing followed by conditioned language synthesis.\looseness=-1

\paragraph{Symbolic Gaze Parser.}
This initial step is a deterministic module that extracts a vocabulary of fundamental gaze events from the raw signal.
For a given segment $i$, the raw gaze signal $S_i^{\text{g}} \in \mathbb{R}^{N_g \times 2}$ consists of a sequence of $N_g$ timestamped coordinates, which we denote as $\{(t_j, y_j, p_j)\}_{j=1}^{N_g}$ where $(y_j, p_j)$ corresponds to the yaw and pitch values. The parser maps this numerical sequence $S_i^{\text{g}}$ into a structured sequence of $M$ symbolic event primitives $E_i = (e_1, e_2, \dots, e_M)$. 
Specifically, we first computed the instantaneous angular velocity $\omega_j$ for each point $j$:
\begin{equation}
\omega_j = \frac{\sqrt{(y_j - y_{j-1})^2 + (p_j - p_{j-1})^2}}{t_j - t_{j-1}}.
\label{eq:i-vt}
\end{equation}
We then follow the established Identification-by-Velocity-Threshold (I-VT) algorithm~\cite{salvucci2000identifying}, which used a two-threshold classification scheme to segment the signal into a vocabulary of primitives, $\mathcal{V} = \{\text{Fixation, Saccade, SmoothPursuit}\}$, based on whether $\omega_j$ falls below a low threshold $v_{\text{low}}$ or exceeds a high threshold $v_{\text{high}}$. 
Each resulting primitive $e_m \in E_i$ is a rich data object, encapsulating not only its class from $\mathcal{V}$ but also quantitative attributes (e.g., duration, amplitude, peak\_velocity) and their corresponding qualitative descriptors (e.g., duration\_label: ``Brief''). This process abstracts the noisy, high-dimensional signal into a compact, machine-readable symbolic representation.\looseness=-1

\paragraph{Symbolic-to-Text Synthesizer.}
The second step leverages an LLM to translate the sequence of gaze symbolic events $E_i$ into a coherent textual narration, $T^{\text{g}}_i$. The module's objective is to generate this narration by modeling the conditional probability $P(T^{\text{g}}_i | E_i)$. We operationalized this by first serializing the sequence of event objects $E_i$ into a descriptive string, which is then embedded within a carefully engineered few-shot prompt~\cite{kojima2022large}. By conditioning the generative process on this structured symbolic input, we fundamentally change the nature of the task for the LLM. Instead of performing risky inference from raw numbers, the model is constrained to a factual translation task: converting a symbolic, verifiable account of behavior into fluent natural language. This ensures the resulting gaze narration is not only descriptive but also verifiably accurate with respect to the underlying physical measurements.

\subsection{Phase 2: Integrating Eye Gaze with Body Motion}\label{sec:integration}
The first phase produces a continuous stream of gaze narrations that remain faithful to the raw signal but are fragmented and lack explanatory depth. Thus, the core challenge is moving from narration to holistic interpretation. Leveraging the strong coupling between gaze and body motion~\cite{hu24pose2gaze}, the second phase integrates gaze narration with atomic body narration to generate a unified, eye–body–coordinated description of human behavior.

\paragraph{Historical Context for Coherence.}
Human behavior exhibits temporal coherence~\cite{wei2022capturing}. To model this dependency, we use a sliding observation window to aggregate the historical context $\mathcal{H}_{i}$. 
Formally, $\mathcal{H}_{i}$ adopts dictionary format,  primarily consisting of two parts: (i) $W$ segments of integrated narrations previously inferred by our pipeline and (ii) feedback content obtained from the last round of self-correcting (Phase 3). Additional scene metadata (e.g., location, focus, etc.) can be included when data is available. $\mathcal{H}_{i}$ is updated in each iteration, efficiently preserving the necessary context and providing a solid foundation for the model to gain insights into temporal coherence and causal relationships.\looseness=-1

\paragraph{Eye-Body-Coordinated Narration Synthesis.}
Then, we can integrate narrations of eye movements and body motions using LLM.
For each segment $i$, the LLM conditions its generation on both the textual gaze narration $T^{\text{g}}_i$ from the first phase of our pipeline and the corresponding sequence of atomic body motion narration $S^{\text{m}}_i$. The objective is to model the conditional probability $P(\hat{T}_i \mid T^{\text{g}}_i, S^{\text{m}}_i, \mathcal{H}_{i})$, where $\mathcal{H}_{i}$ represents the historical context to ensure temporally coherent. 
We first carefully designed a structured prompt template $\Pi_{\text{integ}}(i)$ to fill in all the above information:
\begin{equation}
\Pi_{\text{integ}}(i) = [\texttt{CTX}: \mathcal{H}_{i};\ \texttt{GAZE}: T^{\text{g}}_i;\ \texttt{MOTION}: S^{\text{m}}_i].
\end{equation}
Here, special tokens such as \texttt{CTX:} and \texttt{GAZE:} act as explicit delimiters. This structured formatting is crucial as it guides the LLM to differentiate between historical context, observed gaze behavior, and concurrent body actions, thus facilitating more precise multi-modal reasoning. By conditioning on this structured input, the LLM's task extends beyond mere summarization to active reasoning. For instance, it learns to associate a gaze shift described in $T^{\text{g}}_i$ with a ``user is walking'' description in $S_i^{\text{m}}$ to infer the holistic action: ``The user carefully scans the ground while walking.''

The final integrated eye-body-coordinated narration $\hat{T}_i$ for the current segment is then generated as:
\begin{equation}
\hat{T}_i = \underset{T \in \mathcal{T}}{\arg\max}\ P(T \mid \Pi_{\text{integ}}(i)).
\end{equation}
This mechanism ensures that the resulting narration is not only grounded in the current observation but also logically consistent with preceding actions.

\begin{algorithm}[!t]
\caption{Self-Correcting Loop}
\label{alg:distillation}
\begin{algorithmic}[1]
\Require Initial narration $\hat{T}^{(0)}$, max iterations $K_{\text{max}}$, score thresholds $\boldsymbol{\tau}$
\Ensure A refined, high-quality narration $\hat{T}^{*}$
\State $\hat{T}^{*} \gets \hat{T}^{(0)}$ 
\For{$k = 0$ to $K_{\text{max}}-1$}
    \State $(\mathbf{s}^{(k)}, \phi^{(k)}) \gets \text{LLM}_{\text{eval}}(\hat{T}^{(k)})$ 
    \If{all components $s^{(k)}_j \geq \tau_j$} 
        \State \Return $\hat{T}^{(k)}$ 
    \EndIf
    \State $\hat{T}^{(k+1)} \gets \text{LLM}_{\text{refine}}(\hat{T}^{(k)}, \phi^{(k)})$ 
    \State $\hat{T}^{*} \gets \hat{T}^{(k+1)}$
\EndFor
\State \Return $\hat{T}^{*}$ 
\end{algorithmic}
\end{algorithm}


\begin{table*}[htbp]
    \footnotesize
    \centering
    \setlength{\tabcolsep}{5pt}
    \begin{tabularx}{.98\textwidth}{l|l|XXXXXX}
        \toprule
        Scene Type & Method & \small{MM Dist$\downarrow$} & \small{FID$\downarrow$} & \small{Top-1$\uparrow$} & \small{Top-2$\uparrow$} & \small{Top-3$\uparrow$}  & \small{MM$\uparrow$} \\
        \midrule
        \multirow{2}{*}{Low-level} & MotionGPT & $6.748^{\pm .098}$ & $7.458^{\pm .322}$ & $0.052^{\pm .002}$ & $0.126^{\pm .003}$ & $0.187^{\pm .004}$   & $3.469^{\pm .051}$ \\
        & MotionGPT${\dagger}$& $\mathbf{6.406^{\pm .053}}$ & 
         $\mathbf{6.801^{\pm .241}}$ & 
        $\mathbf{0.102^{\pm .002}}$ & $\mathbf{0.153^{\pm .003}}$ &
         $\mathbf{ 0.214^{\pm .003}}$  &  $\mathbf{3.727^{\pm .044}}$\\
        \midrule
        \multirow{2}{*}{High-level} & MotionGPT & $7.133^{\pm .033}$  & $8.804^{\pm .142}$ & $0.054^{\pm .002}$ & $0.123^{\pm .002}$ & $0.162^{\pm .005}$  & $3.223^{\pm .024}$\\
        & MotionGPT${\dagger}$ & $\mathbf{6.862^{\pm .025}}$ & 
        $\mathbf{8.134^{\pm .323}}$  & 
        $\mathbf{0.062^{\pm .001}}$ & $\mathbf{0.127^{\pm .003}}$ & $\mathbf{0.193^{\pm .004}}$ & 
        $\mathbf{3.864^{\pm .017}}$ \\
        \midrule
        \multirow{2}{*}{All} & MotionGPT & $6.941^{\pm .056}$ & $8.131^{{\pm .304}}$ & $0.053^{\pm .003}$ & $0.124^{\pm .003}$ & $0.175^{\pm .004}$  & $3.346^{{\pm .036}}$ \\
        & MotionGPT${\dagger}$ & $\mathbf{6.634^{\pm .042}}$ & 
        $\mathbf{7.468^{\pm .277}}$ & 
        $\mathbf{0.082^{\pm .004}}$ & $\mathbf{0.140^{\pm .005}}$ & $\mathbf{0.204^{\pm .005}}$ & 
        $\mathbf{3.796^{\pm .028}}$ \\
        \bottomrule
    \end{tabularx}
     \caption{
      Comparison of different input types in text-driven motion generation tasks.
      We fix the generation model weight and evaluate the effect of different text inputs. ${\dagger}$ indicates using our eye-body-coordinated narrations as input.
      }
    \label{tab:text2motion_results}
\end{table*}

\begin{table}[htbp]
    \Huge
    \centering
    \resizebox{\linewidth}{!}{
    \begin{tabular}{l|l|cccc}
        \toprule[1.2pt]
        Type  & Train Set & {Simil.$\uparrow$} & {BERT F1$\uparrow$} & {ROUGE-L$\uparrow$} & {Action F1$\uparrow$}   \\
        \midrule
        \multirow{2}{*}{Low} & Nymeria &$0.525$ &$0.869$ &$0.241$ &$0.281$  \\
        & GazeInterpreter & $\mathbf{0.575}$ & $\mathbf{0.877}$ & $\mathbf{0.276}$ & $\mathbf{0.294}$ \\
        \midrule
        \multirow{2}{*}{High} & Nymeria &$0.393$ &$0.866$ &$0.163$ &$0.171$  \\
        & GazeInterpreter &$\mathbf{0.436}$ &$\mathbf{0.881}$ &$\mathbf{0.186}$ &$\mathbf{0.201}$  \\
        \midrule
        \multirow{2}{*}{All} & Nymeria & $0.459$ & $0.868$ & $0.202$ & $0.226$  \\
        & GazeInterpreter & $\mathbf{0.506}$ & $\mathbf{0.879}$ & $\mathbf{0.231}$ & $\mathbf{0.248}$  \\
        \bottomrule[1.2pt]
    \end{tabular}
    }
     \caption{Comparison of baselines in action anticipation tasks. Simil. indicates Cosine Similarity.}
    \label{tab:action_prediction_results}
\end{table}

\subsection{Phase 3: Self-Correcting Loop}\label{sec:self-correcting}
The output of LLM may fail to emphasize the most salient behavioral cues and suffer from hallucinations~\cite{li2024survey}. We therefore introduce a final self-correcting loop. Unlike generic feedback mechanisms that apply superficial edits~\cite{nacke2011biofeedback}, our loop iteratively evaluates and refines the narration across key quality dimensions, ensuring suitability for downstream tasks requiring a deep understanding of human behavior.

\paragraph{Multi-Dimensional Narration Evaluation.}
The self-correcting loop is driven by a specialized $\text{LLM}_{\text{eval}}$. 
Inspired by~\cite{han2025atom}, we tailor the criteria to the specific dimensions of each narration type.
For gaze narration, the key dimension is Continuity, as natural gaze behavior unfolds as a continuous event stream~\cite{yin2024lg}. This ensures the narration reflects coherent transitions between actions rather than isolated events.
For integrated narration, evaluation considers three factors:
(i)~\textit{Modality Match}, assessing whether gaze and body actions are synergistic and contextually aligned;
(ii)~\textit{Temporal Coherence}, ensuring the narration follows a logical chronological flow; and
(iii)~\textit{Completeness}, verifying that all critical behavioral events are retained.

Based on these dimensions, the output of $\text{LLM}_{\text{eval}}$ is twofold: a structured score vector $\mathbf{s}$ that quantifies quality, and a textual critique $\phi$ that provides targeted feedback for improvement. The specific scoring rubrics that guide this assessment are detailed in Table~\ref{tab:eval_criteria}. This approach ensures the evaluation process is transparent, replicable, and provides fine-grained diagnostics for refinement.

\begin{table}[!t]
    \Huge
    \centering
    \resizebox{\linewidth}{!}{
    \begin{tabular}{l|l|cccc}
        \toprule[1.2pt]
        Type  & Train Set & {Simil.$\uparrow$} & {BERT F1$\uparrow$} & {ROUGE-1$\uparrow$}  & {ROUGE-L$\uparrow$}   \\
        \midrule
        \multirow{2}{*}{Low} & Nymeria & $0.564$ & $0.851$ & $0.228$ & $0.174$  \\
        & \methodName & $\mathbf{0.583}$ & $\mathbf{0.860}$ & $\mathbf{0.283}$ & $\mathbf{0.219}$ \\
        \midrule
        \multirow{2}{*}{High} & Nymeria & $0.395$ & $0.820$ & $0.165$ & $0.126$  \\
        & \methodName &$\mathbf{0.490}$ &$\mathbf{0.859}$ & $\mathbf{0.175}$ & $\mathbf{0.133}$  \\
        \midrule
        \multirow{2}{*}{All} & Nymeria & $0.480$ & $0.836$ & $0.197$ & $0.150$  \\
        & GazeInterpreter & $\mathbf{0.537}$ & $\mathbf{0.860}$ & $\mathbf{0.575}$ & $\mathbf{0.229}$  \\
        \bottomrule[1.2pt]
    \end{tabular}
    }
     \caption{Comparison of baselines in behavior summarization tasks. Simil. indicates Cosine Similarity.
     }
    \label{tab:behavior_summarization_results}
\end{table}

\paragraph{Threshold-Governed Iterative Refinement.}
The refinement process, detailed in Algorithm~\ref{alg:distillation}, is structured as a collaborative loop between two specialized LLMs: $\text{LLM}_{\text{eval}}$ and $\text{LLM}_{\text{refine}}$. The loop iterates until the narration's quality, measured by a score vector $\mathbf{s}$, meets or exceeds a predefined threshold vector $\boldsymbol{\tau}$. In each iteration $k$, $\text{LLM}_{\text{eval}}$ first assesses the current narration $\hat{T}^{(k)}$ to generate both the scores $\mathbf{s}^{(k)}$ and a textual critique $\phi^{(k)}$. If these scores are insufficient, $\text{LLM}_{\text{refine}}$ then uses the original narration and the critique to produce an improved version $\hat{T}^{(k+1)}$. 

This self-correcting loop iteratively filters out noise and redundancy based on set criteria, ensuring that narration correctly reflects facts and ultimately provides a comprehensive overview of human behavior.


\section{Experiments and Results}
Our experimental validation is twofold. We first assess the descriptive fidelity and effectiveness of our integrated narrations via a text-driven motion generation task. Subsequently, we demonstrate their utility for higher-level human behavior understanding on the downstream tasks of action anticipation and behavior summarization.

\begin{figure*}[!t]
    \centering
    \includegraphics[width=\linewidth]{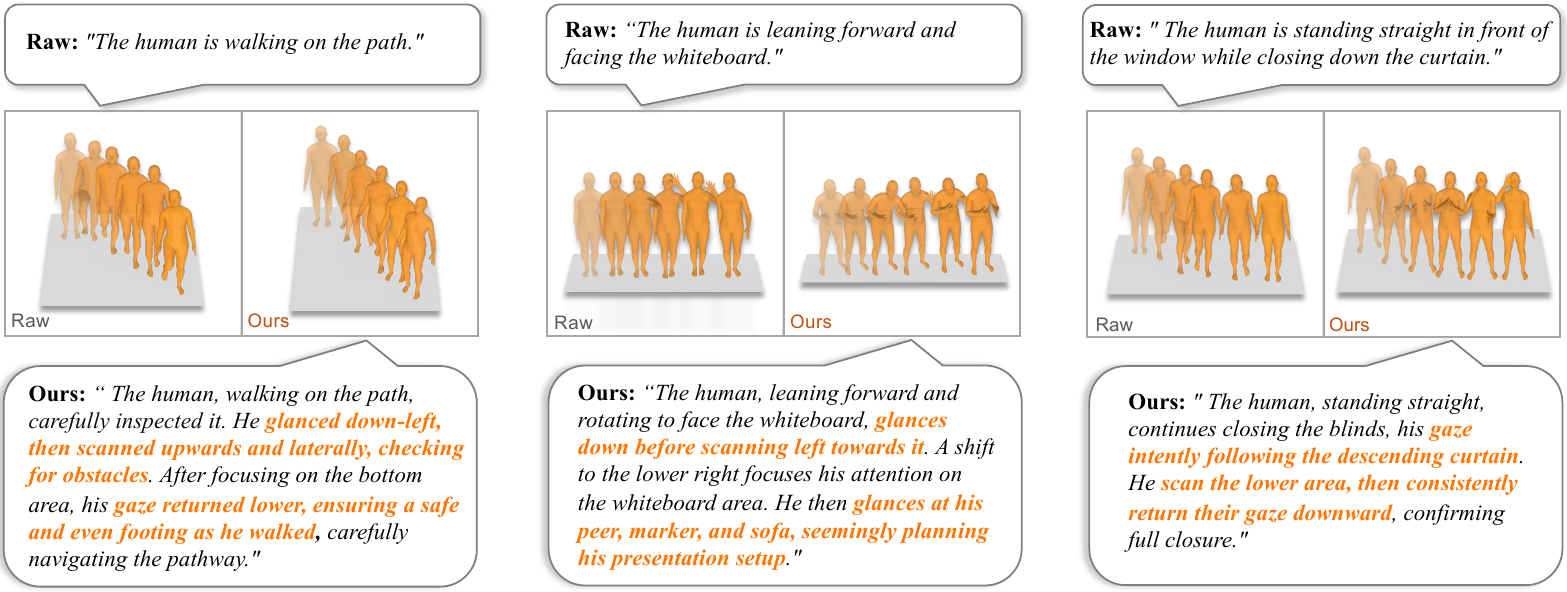}
   \caption{
    Qualitative comparison for text-driven motion generation on \textbf{our} integrated narrations versus \textbf{raw} atomic body narrations, including examples from both low-level scenes (e.g., walking) and high-level scenes (e.g., housework).
}
    \label{fig:vis}
\end{figure*}

\subsection{Implementation Details}
\paragraph{Dataset.}
The Nymeria dataset~\cite{ma2024nymeria} is currently the only public source of synchronized gaze and motion narrations. As the largest human motion dataset with 300 hours of daily activities, its scale allows robust tests. To facilitate a comprehensive analysis of our method's capabilities, we follow the EgoCHARM~\cite{padmanabha2025egocharm} to categorize the 236 sequences with motion narration annotations into two subsets: \textit{high-level} activities, which involve complex, goal-oriented interactions (e.g., housekeeping), and \textit{low-level} activities, which consist of simple, atomic movements (e.g., walking). This division allows us to specifically test our approach's performance on tasks with varying degrees of behavioral complexity.\looseness=-1

\paragraph{Implementation Specifics.}
All LLM-based modules in our framework utilize the Gemini-2.5-Flash model~\cite{comanici2025gemini}, guided by a few-shot, in-context learning strategy~\cite{comanici2025gemini}. For the \textit{symbolic gaze parser}, we follow general guidelines~\cite{salvucci2000identifying} to set the I-VT velocity thresholds to $v_{\text{low}}=30^{\circ}/s$ and $v_{\text{high}}=100^{\circ}/s$. For integrating gaze and body motion, the size of the sliding window is set to $W=2$, and the \textit{self-correcting loop} runs for a maximum of $K_{\text{max}}=3$ iterations with a score threshold of $\tau=4.5$. For more details about LLM prompts and parameters, please refer to our supplementary materials and source code.

\paragraph{Evaluation Metrics.}
We follow the same metrics as~\cite{lin2023motion} to ensure a comprehensive evaluation in the text-driven motion. 
We assess the alignment between text and motion using Multimodal Distance (MM Dist), which measures the average feature space distance, and R-Precision (Top-k), which evaluates retrieval accuracy. 
The realism of the generated motions is measured by the Frechet Inception Distance (FID), which reflects their distributional similarity to real motions. Finally, we evaluate generation diversity using Multimodality (MM), which quantifies the variation among different motions produced from the same text prompt.\looseness=-1

\subsection{Main Experiments}
\paragraph{Experiment Setup.} 
Following Motion-X~\cite{lin2023motion}, we fix the weights of the motion generation model, using the atomic motion narration provided by Nymeria and the integrated narration obtained through our method to drive the model in generating motion sequences, and finally measuring the advantages of our narrations using a wide range of metrics. We choose the popular MotionGPT~\cite{jiang2023motiongpt} model as the motion generation baseline. Moreover, we follow the commonly used feature extractor~\cite{guo2022generating} as the basis for calculating FID metrics.

\paragraph{Quantitative Results.}
When the MotionGPT baseline weights are fixed, Table~\ref{tab:text2motion_results} shows that narrations from \methodName significantly outperform the atomic body narrations provided by Nymeria across all metrics. 
Notably, this advantage is pronounced on low-level activities, where our method substantially reduces the FID from $7.458$ to $6.801$. 
This suggests that the factually-grounded details in our narrations provide a superior conditioning signal for generating precise, atomic motions. The performance gains persist for complex high-level scenes, where \methodName continues to yield lower distribution distances and higher matching scores. 
These findings validate that our eye-body-coordinated narration is effectively grounded in physical behavior and leads to higher-fidelity motion synthesis.

\paragraph{Qualitative Results.}
The qualitative results in Fig.~\ref{fig:vis} vividly demonstrate the superior fidelity of motions generated from our narrations compared to those from atomic body narrations of Nymeria. By capturing the subtle interplay between gaze and physical action, our eye-body-coordinated narration provides crucial contextual richness. This allows the synthesis model to generate actions that are not only more natural and detailed but also more aligned with potential human intentions, providing a solid foundation for a comprehensive explanation of human behavior.

\subsection{Downstream Tasks Experiments}
To demonstrate the advantage of our integrated narrations in comprehensively understanding human behavior, we evaluate their performance on two representative downstream tasks: action anticipation~\cite{hu24hoimotion, hu24gazemotion, hu2024haheae} and behavior summarization~\cite{zhang25summact}. For both tasks, we follow ~\cite{zhang2019bertscore} to employ a comprehensive suite of metrics to assess the generated text. Semantic fidelity is measured by Cosine Similarity and BERTScore F1, while structural coherence and lexical accuracy are evaluated using ROUGE (1/L) and a keyword-based Action F1 Score.

\paragraph{Action Anticipation Results.}
To evaluate the predictive power of our narrations, we perform an action anticipation task where the goal is to generate a textual description of the next action based on the current context. We employ a frozen Gemini-2.5-Flash model as a zero-shot predictor, providing it with either the original Nymeria atomic motion narrations or our integrated narrations as input. As shown in Table~\ref{tab:action_prediction_results}, using the narrations from our method leads to a significant improvement in prediction performance across all metrics. In all cases, our method boosts the cosine similarity of predictions from $0.459$ to $0.506$ and the keyword-centric Action F1 score from $0.226$ to $0.248$. This striking result suggests that by systematically integrating multi-modal cues like gaze, our generated narration encapsulates more predictive, intent-rich information than the human-annotated text alone. This advantage holds for both simple \textit{low-level} and complex \textit{high-level} scenarios, demonstrating the potential for uncovering human intentions.

\paragraph{Behavior Summarization Results.}
This task requires generating a high-level behavior summary from a sequence of motion descriptions. We employed a frozen Gemini-2.5-Flash model, comparing the summaries it produces when conditioned on our integrated narrations versus those from the original Nymeria motion narrations. The results in Table~\ref{tab:behavior_summarization_results} show that our narration serves as a superior conditioning signal, outperforming the baseline across all metrics. This performance gain is especially pronounced in complex \textit{High-level} scenarios, where our method boosts the summary's cosine similarity from $0.395$ to $0.490$. This consistent success across motion generation, anticipation, and summarization tasks compellingly demonstrates that our integrated narrations provide a robust and versatile behavioral representation, effectively enhancing a wide range of human-aware applications.\looseness=-1

\begin{table}    
    \footnotesize
    \centering
     \tabcolsep=0.20cm
    \resizebox{.95\columnwidth}{!}{%
            \begin{tabular}{lcccc}
                \toprule
                Module  & MM Dist $\downarrow$  & FID $\downarrow$  & Top-1 $\uparrow$  \\ 
                 \midrule
               w/o Structure  &$8.135^{\pm .076}$ & $9.124^{\pm .442}$  & $0.059^{\pm .005}$ \\
               w/o Parser   & $7.642^{\pm .060}$ & $7.893^{\pm .459}$   & $0.061^{\pm .005}$ \\
               w/o Loop    & $7.425^{\pm .066}$ & $7.831^{\pm .344}$ & $0.063^{\pm .004}$  \\
               Ours  & $\mathbf{6.634^{\pm .042}}$  & $\mathbf{7.468^{\pm .277}}$  & $\mathbf{0.082^{\pm .004}}$  \\
                \bottomrule
            \end{tabular}}
   \caption{Ablation studies for Hierarchical Structure, Symbolic Gaze Parser, and Self-Correcting Loop.}
    \label{tab:ablation_overall}
\end{table}

\subsection{Ablation Study}
\paragraph{Overall Ablation Study.}
To validate the contribution of our framework's core components, we conduct a comprehensive ablation study, with results presented in Table~\ref{tab:ablation_overall}. The removal of the \textit{hierarchical structure} causes the most significant performance degradation across all metrics, highlighting the importance of integrating gaze semantic extraction into the motion hierarchy. Similarly, ablating the \textit{symbolic gaze parser} by using raw signal inputs, or deactivating the \textit{self-correcting loop}, both lead to a distinct decline in performance. These results validate that each component of the \methodName is essential and contributes effectively to generating high-fidelity behavioral narration.

\paragraph{Ablation Study on Quality Dimension.}
We also analyze the impact of each quality dimension within the self-correcting loop. As detailed in Table~\ref{tab:criteria-ablation}, the results reveal a clear additive effect: cumulatively introducing each of our designed criteria—\textit{Continuity}, \textit{Modality Match}, \textit{Temporal Coherence}, and \textit{Completeness}, progressively improves performance. 
This validates our multi-dimensional evaluation rubric, with each dimension providing an essential criterion for judging how accurately the narrations correspond to real-world behavioral features.

\begin{figure}[!t]
    \centering
    \includegraphics[width=.9\linewidth]{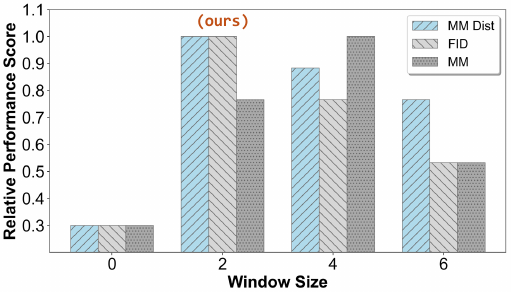} 
   \caption{Effects of different observation window sizes.   
}
    \label{fig:ablation_vis}
\end{figure}

\paragraph{Ablation Study on Windows Size.}
We further analyze the effect of the sliding observation window and investigate how many preceding narration segments $W$ should be incorporated into the historical context $\mathcal{H}_{i}$. As shown in Fig.~\ref{fig:ablation_vis}, a window size of $W=2$ consistently achieves the best balance between contextual richness and computational efficiency. Increasing $W$ provides only marginal gains while noticeably increasing inference cost and occasionally introducing redundant or noisy historical cues. 

\begin{table}[!t]
        \LARGE
	\addtolength{\tabcolsep}{0pt}
	\resizebox{1\columnwidth}{!}{
	\begin{tabular}{*{7}{c}}
		\toprule[1.2pt]
		Continuity & Match & Temporal & Completeness   & Top-1$\uparrow$  & FID$\downarrow$ \\
		\midrule
                          &                          &                  &   & 0.063 &  7.831    \\
		\checkmark            &                          &                  &   & 0.069 &  7.722    \\
		\checkmark            & \checkmark               &                  &   & 0.072  & 7.644  \\
        \checkmark            & \checkmark               &   \checkmark                &   & 0.074  & 7.573  \\
        
		\checkmark            & \checkmark               & \checkmark       &  \checkmark & \textbf{0.082} & \textbf{7.468}  \\
		\bottomrule[1.2pt]
	\end{tabular}
	}
        \caption{Effects of each evaluation dimension.}
        \label{tab:criteria-ablation}
\end{table}

\section{Conclusion}
In this work, we proposed a novel LLM-based method that features a symbolic gaze parser to convert raw gaze signal, a hierarchical framework to integrate gaze with body motion, and a self-correcting loop for refinement.
Through extensive experiments on a large-scale benchmark, we showed the advantages of our eye-body-coordinated narrations in text-driven motion generation.
We also demonstrated the effectiveness of our method for the sample downstream tasks of action anticipation and behavior summarization.
As such, our work reveals the significant information content
available in eye gaze for interpreting human behavior and guides future work on this promising direction.

\clearpage
\newpage
\bibliography{aaai2026}

@ARTICLE{sitzmann2018saliency, 
author={Sitzmann, Vincent and Serrano, Ana and Pavel, Amy and Agrawala, Maneesh and Gutierrez, Diego and Masia, Belen and Wetzstein, Gordon}, 
journal={IEEE Transactions on Visualization and Computer Graphics}, 
title={Saliency in VR: how do people explore virtual environments?}, 
year={2018}, 
volume={24}, 
number={4}, 
pages={1633-1642}
}

@inproceedings{zhang25summact,
	title = {SummAct: Uncovering User Intentions Through Interactive Behaviour Summarisation},
	author = {Zhang, Guanhua and Ahmed, Mohamed and Hu, Zhiming and Bulling, Andreas},
	year = {2025},
	pages = {1--17},
	booktitle = {Proc. ACM CHI Conference on Human Factors in Computing Systems (CHI)}}

@inproceedings{nacke2011biofeedback,
  title={Biofeedback game design: using direct and indirect physiological control to enhance game interaction},
  author={Nacke, Lennart Erik and Kalyn, Michael and Lough, Calvin and Mandryk, Regan Lee},
  booktitle={Proceedings of the SIGCHI conference on human factors in computing systems},
  pages={103--112},
  year={2011}
}

@article{hadnett2019effect,
  title={The effect of task on visual attention in interactive virtual environments},
  author={Hadnett-Hunter, Jacob and Nicolaou, George and O'Neill, Eamonn and Proulx, Michael},
  journal={ACM Transactions on Applied Perception},
  volume={16},
  number={3},
  pages={1--17},
  year={2019},
  publisher={ACM New York, NY, USA}
}

@inproceedings{hu25hoigaze,
	title={HOIGaze: Gaze Estimation During Hand-Object Interactions in Extended Reality Exploiting Eye-Hand-Head Coordination},
	author={Hu, Zhiming and Haeufle, Daniel and Schmitt, Syn and Bulling, Andreas},
	booktitle={Proceedings of the ACM Special Interest Group on Computer Graphics and Interactive Techniques},
	year={2025},
	pages = {1--10}}

@article{hu2022ehtask,
  title = {EHTask: recognizing user tasks from eye and head movements in immersive virtual reality},
  author = {Hu, Zhiming and Bulling, Andreas and Li, Sheng and Wang, Guoping},
  year = {2022},
  journal = {IEEE Transactions on Visualization and Computer Graphics}
}

@article{hu2019sgaze,
  title={SGaze: a data-driven eye-head coordination model for realtime gaze prediction},
  author={Hu, Zhiming and Zhang, Congyi and Li, Sheng and Wang, Guoping and Manocha, Dinesh},
  journal={IEEE Transactions on Visualization and Computer Graphics},
  volume={25},
  number={5},
  pages={2002--2010},
  year={2019},
  publisher={IEEE}
}

@article{hu2020dgaze,
  title={DGaze: CNN-based gaze prediction in dynamic scenes},
  author={Hu, Zhiming and Li, Sheng and Zhang, Congyi and Yi, Kangrui and Wang, Guoping and Manocha, Dinesh},
  journal={IEEE Transactions on Visualization and Computer Graphics},
  volume={26},
  number={5},
  pages={1902--1911},
  year={2020},
  publisher={IEEE}
}

@article{hu2021fixationnet,
  title={FixationNet: forecasting eye fixations in task-oriented virtual environments},
  author={Hu, Zhiming and Bulling, Andreas and Li, Sheng and Wang, Guoping},
  journal={IEEE Transactions on Visualization and Computer Graphics},
  volume={27},
  number={5},
  pages={2681--2690},
  year={2021},
  publisher={IEEE}
}

@inproceedings{pons2014posebits,
  title={Posebits for monocular human pose estimation},
  author={Pons-Moll, Gerard and Fleet, David J and Rosenhahn, Bodo},
  booktitle={Proceedings of the IEEE Conference on Computer Vision and Pattern Recognition},
  pages={2337--2344},
  year={2014}
}

@inproceedings{chen2024gazexplain,
  title={Gazexplain: Learning to predict natural language explanations of visual scanpaths},
  author={Chen, Xianyu and Jiang, Ming and Zhao, Qi},
  booktitle={European Conference on Computer Vision},
  pages={314--333},
  year={2024},
  organization={Springer}
}

@inproceedings{delmas2022posescript,
  title={Posescript: 3d human poses from natural language},
  author={Delmas, Ginger and Weinzaepfel, Philippe and Lucas, Thomas and Moreno-Noguer, Francesc and Rogez, Gr{\'e}gory},
  booktitle={European Conference on Computer Vision},
  pages={346--362},
  year={2022},
  organization={Springer}
}

@article{sidenmark2019eye,
  title={Eye, head and torso coordination during gaze shifts in virtual reality},
  author={Sidenmark, Ludwig and Gellersen, Hans},
  journal={ACM Transactions on Computer-Human Interaction},
  volume={27},
  number={1},
  pages={1--40},
  year={2019},
  publisher={ACM New York, NY, USA}
}

@article{itti1998model,
  title={A model of saliency-based visual attention for rapid scene analysis},
  author={Itti, Laurent and Koch, Christof and Niebur, Ernst},
  journal={IEEE Transactions on Pattern Analysis and Machine Intelligence},
  volume={20},
  number={11},
  pages={1254--1259},
  year={1998},
  publisher={IEEE}
}

@book{yarbus1967eye,
  title={Eye movements and vision},
  author={Yarbus, Alfred L},
  publisher={Springer},
  year={1967}
}

@inproceedings{salvucci2000identifying,
  title={Identifying fixations and saccades in eye-tracking protocols},
  author={Salvucci, Dario D and Goldberg, Joseph H},
  booktitle={Proceedings of the 2000 ACM Symposium on Eye Tracking Research and Applications},
  pages={71--78},
  year={2000}
}

@article{coutrot2018scanpath,
  title={Scanpath modeling and classification with hidden Markov models},
  author={Coutrot, Antoine and Hsiao, Janet H and Chan, Antoni B},
  journal={Behavior Research Methods},
  volume={50},
  number={1},
  pages={362--379},
  year={2018},
  publisher={Springer}
}

@article{kothari2020gaze,
  title={Gaze-in-wild: a dataset for studying eye and head coordination in everyday activities},
  author={Kothari, Rakshit and Yang, Zhizhuo and Kanan, Christopher and Bailey, Reynold and Pelz, Jeff B and Diaz, Gabriel J},
  journal={Scientific Reports},
  volume={10},
  number={1},
  pages={1--18},
  year={2020},
  publisher={Nature Publishing Group}
}

@article{hu24pose2gaze,
	author={Hu, Zhiming and Xu, Jiahui and Schmitt, Syn and Bulling, Andreas},
	journal={IEEE Transactions on Visualization and Computer Graphics}, 
	title={Pose2Gaze: Eye-body Coordination during Daily Activities for Gaze Prediction from Full-body Poses}, 
	year={2024}}

@article{jiang2023motiongpt,
  title={Motiongpt: Human motion as a foreign language},
  author={Jiang, Biao and Chen, Xin and Liu, Wen and Yu, Jingyi and Yu, Gang and Chen, Tao},
  journal={Advances in Neural Information Processing Systems},
  volume={36},
  pages={20067--20079},
  year={2023}
}

@article{chen2024motionllm,
  title={Motionllm: Understanding human behaviors from human motions and videos},
  author={Chen, Ling-Hao and Lu, Shunlin and Zeng, Ailing and Zhang, Hao and Wang, Benyou and Zhang, Ruimao and Zhang, Lei},
  journal={arXiv preprint arXiv:2405.20340},
  year={2024}
}

@inproceedings{belardinelli2022intention,
  title={Intention estimation from gaze and motion features for human-robot shared-control object manipulation},
  author={Belardinelli, Anna and Kondapally, Anirudh Reddy and Ruiken, Dirk and Tanneberg, Daniel and Watabe, Tomoki},
  booktitle={Proceedings of the 2022 IEEE International Conference on Intelligent Robots and Systems},
  pages={9806--9813},
  year={2022},
  organization={IEEE}
}

@article{hu2024haheae,
  title={HaHeAE: Learning Generalisable Joint Representations of Human Hand and Head Movements in Extended Reality},
  author={Hu, Zhiming and Zhang, Guanhua and Yin, Zheming and Haeufle, Daniel and Schmitt, Syn and Bulling, Andreas},
  journal={arXiv preprint arXiv:2410.16430},
  year={2024}
}

@article{hu24hoimotion,
	author={Hu, Zhiming and Yin, Zheming and Haeufle, Daniel and Schmitt, Syn and Bulling, Andreas},
	journal={IEEE Transactions on Visualization and Computer Graphics}, 
	title={HOIMotion: Forecasting Human Motion During Human-Object Interactions Using Egocentric 3D Object Bounding Boxes}, 
	year={2024}}

@article{zhang2022motiondiffuse,
  title={Motiondiffuse: Text-driven human motion generation with diffusion model},
  author={Zhang, Mingyuan and Cai, Zhongang and Pan, Liang and Hong, Fangzhou and Guo, Xinying and Yang, Lei and Liu, Ziwei},
  journal={arXiv preprint arXiv:2208.15001},
  year={2022}
}

@inproceedings{yan24gazemodiff,
	title={GazeMoDiff: Gaze-guided Diffusion Model for Stochastic Human Motion Prediction},
	author={Yan, Haodong and Hu, Zhiming and Schmitt, Syn and Bulling, Andreas},
	booktitle={Proceedings of the 2024 Pacific Conference on Computer Graphics and Applications},	
	year={2024}}

@inproceedings{hu24gazemotion,
	title={GazeMotion: Gaze-guided Human Motion Forecasting},
	author={Hu, Zhiming and Schmitt, Syn and Haeufle, Daniel and Bulling, Andreas},
	booktitle={Proceedings of the 2024 IEEE/RSJ International Conference on Intelligent Robots and Systems},	
	year={2024}}

@article{bommasani2021opportunities,
  title={On the opportunities and risks of foundation models},
  author={Bommasani, Rishi and Hudson, Drew A and Adeli, Ehsan and Altman, Russ and Arora, Simran and von Arx, Sydney and Bernstein, Michael S and Bohg, Jeannette and Bosselut, Antoine and Brunskill, Emma and others},
  journal={arXiv preprint arXiv:2108.07258},
  year={2021}
}

@article{lee2023well,
  title={How well do large language models truly ground?},
  author={Lee, Hyunji and Joo, Sejune and Kim, Chaeeun and Jang, Joel and Kim, Doyoung and On, Kyoung-Woon and Seo, Minjoon},
  journal={arXiv preprint arXiv:2311.09069},
  year={2023}
}

@article{wang2024can,
  title={Can LLMs Convert Graphs to Text-Attributed Graphs?},
  author={Wang, Zehong and Liu, Sidney and Zhang, Zheyuan and Ma, Tianyi and Zhang, Chuxu and Ye, Yanfang},
  journal={arXiv preprint arXiv:2412.10136},
  year={2024}
}

@inproceedings{ma2024nymeria,
  title={Nymeria: A massive collection of multimodal egocentric daily motion in the wild},
  author={Ma, Lingni and Ye, Yuting and Hong, Fangzhou and Guzov, Vladimir and Jiang, Yifeng and Postyeni, Rowan and Pesqueira, Luis and Gamino, Alexander and Baiyya, Vijay and Kim, Hyo Jin and others},
  booktitle={European Conference on Computer Vision},
  pages={445--465},
  year={2024},
  organization={Springer}
}

@article{padmanabha2025egocharm,
  title={EgoCHARM: Resource-Efficient Hierarchical Activity Recognition using an Egocentric IMU Sensor},
  author={Padmanabha, Akhil and Govindarajan, Saravanan and Kim, Hwanmun and Ortiz, Sergio and Rajan, Rahul and Senkal, Doruk and Kadetotad, Sneha},
  journal={arXiv preprint arXiv:2504.17735},
  year={2025}
}

@article{comanici2025gemini,
  title={Gemini 2.5: Pushing the frontier with advanced reasoning, multimodality, long context, and next generation agentic capabilities},
  author={Comanici, Gheorghe and Bieber, Eric and Schaekermann, Mike and Pasupat, Ice and Sachdeva, Noveen and Dhillon, Inderjit and Blistein, Marcel and Ram, Ori and Zhang, Dan and Rosen, Evan and others},
  journal={arXiv preprint arXiv:2507.06261},
  year={2025}
}

@article{lin2023motion,
  title={Motion-x: A large-scale 3d expressive whole-body human motion dataset},
  author={Lin, Jing and Zeng, Ailing and Lu, Shunlin and Cai, Yuanhao and Zhang, Ruimao and Wang, Haoqian and Zhang, Lei},
  journal={Advances in Neural Information Processing Systems},
  volume={36},
  pages={25268--25280},
  year={2023}
}

@inproceedings{han2025atom,
  title={Atom: Aligning text-to-motion model at event-level with gpt-4vision reward},
  author={Han, Haonan and Wu, Xiangzuo and Liao, Huan and Xu, Zunnan and Hu, Zhongyuan and Li, Ronghui and Zhang, Yachao and Li, Xiu},
  booktitle={Proceedings of the Computer Vision and Pattern Recognition Conference},
  pages={22746--22755},
  year={2025}
}

@article{kojima2022large,
  title={Large language models are zero-shot reasoners},
  author={Kojima, Takeshi and Gu, Shixiang Shane and Reid, Machel and Matsuo, Yutaka and Iwasawa, Yusuke},
  journal={Advances in neural information processing systems},
  volume={35},
  pages={22199--22213},
  year={2022}
}

@inproceedings{wei2022capturing,
  title={Capturing humans in motion: Temporal-attentive 3D human pose and shape estimation from monocular video},
  author={Wei, Wen-Li and Lin, Jen-Chun and Liu, Tyng-Luh and Liao, Hong-Yuan Mark},
  booktitle={Proceedings of the IEEE/CVF Conference on Computer Vision and Pattern Recognition},
  pages={13211--13220},
  year={2022}
}

@article{yan2023efficient,
  title={An efficient end-to-end training approach for zero-shot human-AI coordination},
  author={Yan, Xue and Guo, Jiaxian and Lou, Xingzhou and Wang, Jun and Zhang, Haifeng and Du, Yali},
  journal={Advances in neural information processing systems},
  volume={36},
  pages={2636--2658},
  year={2023}
}

@article{xu2025distinguishing,
  title={Distinguishing llm-generated from human-written code by contrastive learning},
  author={Xu, Xiaodan and Ni, Chao and Guo, Xinrong and Liu, Shaoxuan and Wang, Xiaoya and Liu, Kui and Yang, Xiaohu},
  journal={ACM Transactions on Software Engineering and Methodology},
  volume={34},
  number={4},
  pages={1--31},
  year={2025},
  publisher={ACM New York, NY}
}

@inproceedings{yin2024lg,
  title={Lg-gaze: Learning geometry-aware continuous prompts for language-guided gaze estimation},
  author={Yin, Pengwei and Wang, Jingjing and Zeng, Guanzhong and Xie, Di and Zhu, Jiang},
  booktitle={European Conference on Computer Vision},
  pages={1--17},
  year={2024},
  organization={Springer}
}

@article{zhang2019bertscore,
  title={Bertscore: Evaluating text generation with bert},
  author={Zhang, Tianyi and Kishore, Varsha and Wu, Felix and Weinberger, Kilian Q and Artzi, Yoav},
  journal={arXiv preprint arXiv:1904.09675},
  year={2019}
}

@inproceedings{chang2024hybrid,
  title={A hybrid global-local perception network for lane detection},
  author={Chang, Qing and Tong, Yifei},
  booktitle={Proceedings of the AAAI Conference on Artificial Intelligence},
  volume={38},
  pages={981--989},
  year={2024}
}

@inproceedings{chang2025spatial,
  title={Spatial-Temporal Perception with Causal Inference for Naturalistic Driving Action Recognition},
  author={Chang, Qing and Dai, Wei and Shuai, Zhihao and Yu, Limin and Yue, Yutao},
  booktitle={ICASSP 2025-2025 IEEE International Conference on Acoustics, Speech and Signal Processing (ICASSP)},
  pages={1--5},
  year={2025},
  organization={IEEE}
}

@article{kong2022human,
  title={Human action recognition and prediction: A survey},
  author={Kong, Yu and Fu, Yun},
  journal={International Journal of Computer Vision},
  volume={130},
  number={5},
  pages={1366--1401},
  year={2022},
  publisher={Springer}
}

@article{li2024survey,
  title={A survey on LLM-based multi-agent systems: workflow, infrastructure, and challenges},
  author={Li, Xinyi and Wang, Sai and Zeng, Siqi and Wu, Yu and Yang, Yi},
  journal={Vicinagearth},
  volume={1},
  number={1},
  pages={9},
  year={2024},
  publisher={Springer}
}

@inproceedings{guo2022generating,
  title={Generating diverse and natural 3d human motions from text},
  author={Guo, Chuan and Zou, Shihao and Zuo, Xinxin and Wang, Sen and Ji, Wei and Li, Xingyu and Cheng, Li},
  booktitle={Proceedings of the IEEE/CVF conference on computer vision and pattern recognition},
  pages={5152--5161},
  year={2022}
}

\end{document}